**Sputtered-silica defect layer in artificial opals: tunability of highly transmitted and reflected optical modes**

*Phan Ngoc Hong*, *Paul Benalloul*, *Laurent Coolen*, *Agnès Maître*, and *Catherine Schwob*[*]

Phan Ngoc Hong, Dr. Paul Benalloul, Dr. Laurent Coolen, Prof. Agnès Maître, Prof. Catherine Schwob
Institut des NanoSciences de Paris, University Pierre et Marie Curie, CNRS, UMR 7588
4 place Jussieu, Paris 75252 cedex 05, France
E-mail: (schwob@insp.jussieu.fr)
Phan Ngoc Hong
Institute of Materials Science, Vietnam Academy of Science and Technology
18 Hoang Quoc Viet road, Cau Giay distr., Hanoi, Vietnam

We propose an original and efficient method to engineer a defect between two well-ordered silica opals by sputtering silica on the top of the first one. As the amount of sputtered silica can be well controlled, it is also the case for the thickness of the layer and consequently for the spectral position of the defect mode. The optical response of these sandwich structures is studied in terms of specular reflection and transmission spectroscopy. Tunable highly transmitted and reflected optical modes are evidenced. The very good agreement between the experimental results and the simulations, run without fitting parameters, demonstrates the almost perfect order of the synthesized structures.

Artificial opals are 3D photonic crystals, the synthesis of which is based on self-assembly of dielectric spheres.[1,2] They are objects of great interest since the last decades as they can allow light confinement in the three directions of space (in the case where they exhibit complete photonic bandgap) and as their fabrication methods are cost efficient, versatile and reliable.[1,3]

3D-photonic crystals find applications in low-threshold lasers,[4] low-loss waveguides,[5,6,7] on-chip optical circuits,[8] fiber optics[9,10] and for the control of the fluorescence properties of quantum dots[11,12,13] or other fluorophores.[14,15,16] Further tailoring the photonic properties is provided by embedding a controlled defect in the structure. Indeed, the disruption of the photonic crystal periodicity can create permitted optical frequency bands within the photonic stopband: light frequency included in the corresponding passband is then localized in the defect allowing waveguiding or optical cavity effects for example. Several fabrication methods, reviewed in,[17] have been used to engineer defect layers sandwiched between two opals.

The first defect layers were created using the Langmuir-Blodgett technique as it is an efficient way to transfer monolayers of beads on a substrate (which may be an opal). The defect layer can be obtained with spheres of different diameter with respect to the ones which compose the opals.[18,19,20] Although it gave interesting results in bandgap engineering, the main drawback of this method is that it creates disorder in the arrangement of the second opal, due to the size difference of the periodic structures (defect layer and second opal). Indeed, it leads to dips of weak amplitude in the experimental reflection spectra. More recently, defect layers with spheres of same size but different index were embedded in opals.[21] Up to now, the corresponding samples show very unpronounced dips compared to simulation predictions. This strategy is inherently limited by the fact that tuning the defect mode requires tuning the spheres index, which is very difficult.

To create a planar defect, techniques based on spin-coating of nanocrystalline colloids such as $TiO_2$[22,23] or on polyelectrolyte multilayers[24] have been proposed and demonstrated.

Both methods lead to very pronounced dips in the reflectivity spectrum, but high transmission may be limited by the scattering of the nanocrystallites.

Some groups have reported the design of a luminescent plane defect layer. In reference,[25] the plasma-enhanced chemical vapor deposition technique is used to create a luminescent organic defect layer consisting in a close-packed arrangement of cylindrical nanopillars oriented perpendicularly to the opal surface. In this case, the dip in the reflectivity spectrum represents about 20% of the maximum.

Another strategy proposed to embed a defect layer between two inverse opals.[26, 27] In reference,[26] reflection and transmission experimental spectra evidence a high contrast between reflected and transmitted modes. In reference[27] a good agreement between the experimental positions of the defect mode and the corresponding simulations while varying the defect thickness is shown.

In this paper, we propose an original and efficient method to engineer a defect between two well-ordered direct silica opals by sputtering silica on the top of the first one. The upper opal has a high crystallinity which is moreover connected to the one of the lower opal. Our experimental data evidence a very high contrast (almost 100%) between reflected and transmitted modes, demonstrating that our samples are well-ordered and present low light scattering. Their geometry is well controlled as evidenced by the almost perfect agreement between experiments and simulations.

The contents are organized as follow: after the presentation of the synthesis protocol and the structural characterization, we will study the optical properties of one particular sample by specular reflection spectroscopy. We will then present simulations of light propagation in the structures and compare the results to experimental transmission spectra. Finally, we will study the spectral position of the transmitted mode with respect to the thickness of sputtered silica layer and confront the experimental results to the simulations.

We synthesized silica spheres by a procedure derived from the Stöber-Fink-Bohn technique.[28] We measured their diameter and size dispersion by SEM images of the opals synthesized from these spheres: the mean diameter was 363 nm and the size dispersion was on the order of 3.5%.

A first opal was fabricated by a convection self-assembly protocol similar to reference:[29] silica beads diluted in ethanol with a concentration of 1.25 wt% self-organized on a glass substrate 10°-tilted from the vertical with a 5°C temperature gradient between the bottom (25°C) and the top (20°C) of the vessel. The thickness of the opal was measured by SEM. For the sample corresponding to the **Figure 3** and **Figure 6**, the thickness was found to be equal to (2.87 ±0.03) µm; it corresponded to 9 monolayers. The thickness of the other samples varied of ± 2 layers with respect to the latter. Then silica, with an optical index of 1.49 (deduced from ellipsometric measurements made on a reference silica layer deposited on a silicon substrate), was sputtered on the top of the opal from a $SiO_2$ target under argon atmosphere. The sample surface was oriented to be parallel to the target so that the sputtering took place with a preferred direction, perpendicular to the sample surface ((111) planes). The parameters of sputtering (pressure of the chamber, valve opening) were calibrated through silica deposition on a reference substrate before deposition on the opal. It is well-established that thin films do not grow uniformly on bulk substrates.[30] In sputtering process, the growth direction mainly depends on the gas pressure and on the ratio between substrate temperature and fusion temperature of the material. In our case (argon pressure equal to 0.2 Pa and relative temperature equal to 0.2), we obtained a structure made of columns. These columns grew on the top of the opal upper layer spheres and reproduced well the opal periodicity. This effect appears clearly on SEM images (**Figure 1**): the sputtered silica covers the top of the spheres of the opal upper layer, leading to a close-packed arrangement of oriented elongated beads. The thickness of deposited silica, named *e* in the following, is defined as the width of the cap covering the silica spheres (**Figure 5**). It was estimated by SEM images as the difference between the height of the elongated beads and the opal-beads diameter. We synthesized samples of different thicknesses *e* varying from 41 to 273 nm, measured with an uncertainty of ±6 nm. These thicknesses were chosen, after running simulations, to produce defect modes on the edge (extreme thicknesses samples: *e*=41 nm and *e*=273 nm) and on the middle (median thicknesses samples: *e*=87 nm and *e*=123 nm) of the stopband.

Finally, the sputtered silica was covered by a second opal, obtained by the same protocol as the first one. The thickness of the second opal was 3.03 µm (10 monolayers).

The final sandwich structure is shown on **Figure 1**. As the close-packed arrangement of the nanocylinders reproduces the one of the first opal, there is not any disrupt of the initial periodicity in the (111) planes (given by the sample surface). The defect layer surface constitutes a desirable template for deposition of the second opal. Consequently, the final sandwich structure demonstrates a well-preserved 3D-order.

We performed specular reflection spectroscopy on our structures. The spectra were measured at various incidence angles θ, defined from the normal to the (111) crystallographic planes. The incidence beam was provided by a halogen lamp connected to an optical fiber (core 600 $\mu$m), mounted on a goniometer arm with a collimator (focal length 12.7mm) and a diaphragm (diameter 0.6 mm). The reflected beam was collected by a symmetric collimated fiber. Due to the beam divergence (1.4°), the beam diameter is larger than 3 mm at the entrance of the detection fiber. To ensure the detection of specular reflection with a resolution of 0.5°, we placed a 1-mm diaphragm at the entrance of this fiber. The detected signal was analyzed by a spectrometer (resolution Δλ=1.5 nm). The transmission spectra were performed with the same setup with light propagation and detection directions normal to the sample plane. In order to take the setup response into account, all the spectra were normalized by the source spectrum.

Note that the incident light was unpolarized so that the detected signal included both s and p-polarizations.

First, we characterized the first opal sample (9 layers) covered by 123 nm of sputtered silica by specular reflectivity (**Figure 2**). When the incidence angle θ increased, the reflectivity peak, and so the stopband, appeared for shorter wavelengths, showing the photonic crystal effect. The reflection coefficient varied between 17% and 33% depending on the incidence angle as already described in[31] for instance. The full width at half-maximum of the peak was of the order of 60 nm. For θ larger than 40°, a second reflection maximum was observed. This peak, due to diffraction on other crystallographic planes (200 and -111), testified of the good 3D-order of the opal. The value of the opal effective index $n_{eff}$ was deduced from the method, based on the reflectivity spectra, developed in.[31] From the obtained value $n_{eff}$=1.25 and the assumption that the opal structure was fcc (eg filling factor $f$=0.74), the optical index of the silica spheres, $n_1$= 1.32, was deduced from the relationship $n_{eff}^2 = f n_1^2 + (1-f)$. This result evidenced the porosity of the synthesized silica beads as already studied in.[32] It should be noticed that the silica layer has not modified the photonic crystal character of the opal.

Let us now focus on the reflection spectra of the sandwich structure with the 123 nm-sputtered silica layer. We performed measurements for specular angles between 20° and 50° by step of 5°. **Figure 3** shows the reflection spectra for 20° and 35° incidence angles. It appeared clearly that, thanks to the defect, a passband was created in the stopband. The spectral width at half maximum of this passband was on the order of 25 nm. The value of the reflection minimum was almost zero. The corresponding contrast between reflected and transmitted optical modes, on the order of 90%, is very high compared to the best results previously reported in the literature.

We plotted on **Figure 4** the spectral positions of the reflection minima for different incidence angles (circles on **Figure 4**). The two extreme curves correspond to the limits of the stopband taken to be the wavelengths closest to the maximum for which the reflection goes to zero on each side (see arrows on fig 3 for 20°-angle).

By varing the angle from 20° to 50°, the defect mode spectral position goes from 717 nm to 598 nm, leading to a high tunability of almost 120 nm. Consequently, this kind of sandwich structure should be very suitable to allow wavelength-selective excitation and so to address selectively different emitters.

Simulations of light propagation through these structures have been performed with the Finite Difference Time Domain (FDTD) method, using the freely available software MEEP developed in MIT. [33]

The computed 3D close-packed structure is shown on **Figure 5**. For the first and second opals, the beads diameter and silica index values, deduced from structural and optical characterizations of our samples, were set respectively to $D_1$=363 nm and $n_1$=1.32. To faithfully reproduce the shape of the defect, a monolayer of beads of same diameter $D_1$ and of index $n_2$=1.49 (index of the sputtered silica from ellipsometry) was inserted between the two opals. The position of the beads of index $n_2$ was set so that the distance between the top of these beads and the top of the first opal upper layer was equal to the thickness of sputtered silica measured by SEM (inset of **Figure 5**). Moreover, to fill the voids between the two kinds of spheres, a rectangular layer of index $n_2$ and of width *e* is added between the centers of the two layers of beads.

We simulated the 0°-transmission spectra of our samples illuminated by a light source of constant intensity over the visible range and compared them with the experimental ones. **Figure 6** shows the experimental and calculated transmission spectra of the 123 nm-defect thickness sample.

We obtained a very good agreement between the calculated and measured spectral positions of the defect mode, emphasizing the fact that our computed structure, simulated without any fitting parameter, is relevant. It allows us to predict its position as a function of the thickness of sputtered silica. As an example, for the 123 nm-defect thickness sample, the discrepancy between simulation and experiment for the position of the defect mode is as low as $\Delta\lambda=5$ nm. Finally, we studied the position of the defect mode for various thicknesses of sputtered silica: $e$=41 nm; 87 nm; 123 nm and 273 nm. We measured the corresponding transmission spectra and performed the simulations. Simulations were also ran for others thicknesses between 150 and 400 nm. The results were summarized on **Figure 8** which gives the spectral position of the defect mode as a function of the sputtered silica thickness. The experimental data are in very good agreement with the simulations except for the case $e$=273 nm for which the relief of the defect is smoothed by the high amount of deposited silica. For this sample, the rectangular layer of index $n_2$ in the computed structure should have a width larger than $e$ to reproduce faithfully the shape of the defect. Indeed, we increased this width up to 450 nm and obtained a very good agreement between experiment and simulation (the corresponding simulation is not shown on **Figure 8**). By controlling the amount of sputtered silica, we were able to monitor the position of the defect mode from the edge to the middle of the stopband (**Figure 7**).

To conclude, we developed a very efficient and reliable experimental method to engineer a defect layer between two artificial opals. Thanks to the growth direction of the sputtered silica, we were able to preserve the periodicity of the sample in the (111) direction and so to obtain a good 3D-order of the sandwich structure. As shown by the very good agreement between simulated and experimental transmission spectra, our samples are very close to perfectly-ordered structures.

Thanks to an accurate design of the defect structure in FDTD simulations, we were able to precisely predict the position of the defect optical mode with respect to the photonic stopband and consequently to fabricate samples with a passband located in the middle of the stopband.

The experimental reflection spectra evidenced a nearly 100%-contrast between reflected and transmitted modes and a high tunability of their spectral position obtained by varying the incidence angle. Consequently, these structures are of great interest to manipulate the optical properties of photonic crystals.

**Acknowledgements**

(The authors would like to thank Eric Charron and Willy Daney de Marcillac (INSP) for their work on the goniometer setup, Dominique Demaille (INSP) for her help on SEM measurements, Stéphane Chenot (INSP) for silica sputtering and Carlos Barthou (INSP) for its fruitful advice. They are grateful to Kifle Aregahegn and Juan-Sebastian Restrepo for their contributions, during their intership in INSP, respectively on silica beads and opals synthesis and on simulations.

The collaboration between INSP and IMS was supported by a Projet International de Coopération Scientifique (PICS 5724) between CNRS and VAST.)

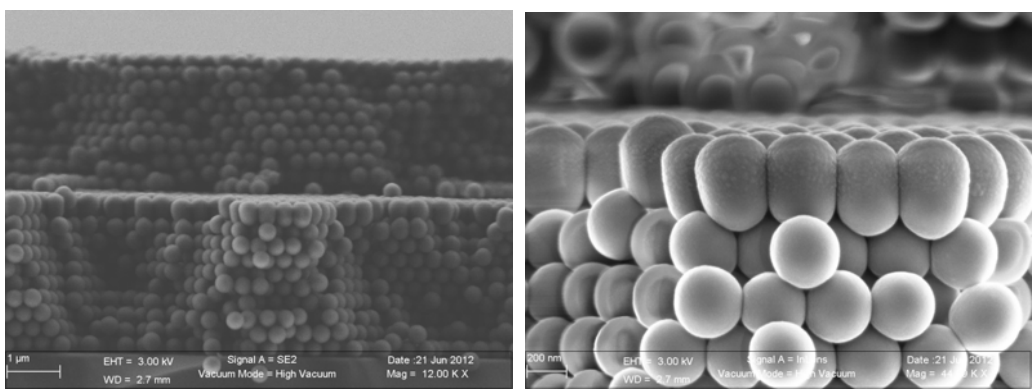

**Figure 1.** (SEM images of a structure: a layer of sputtered silica (silica index 1.49, thickness 123nm) is sandwiched between two opals (silica index 1.32, spheres mean diameter 363 nm). The picture on the right is a magnification of the sputtered silica layer and of the first opal.)

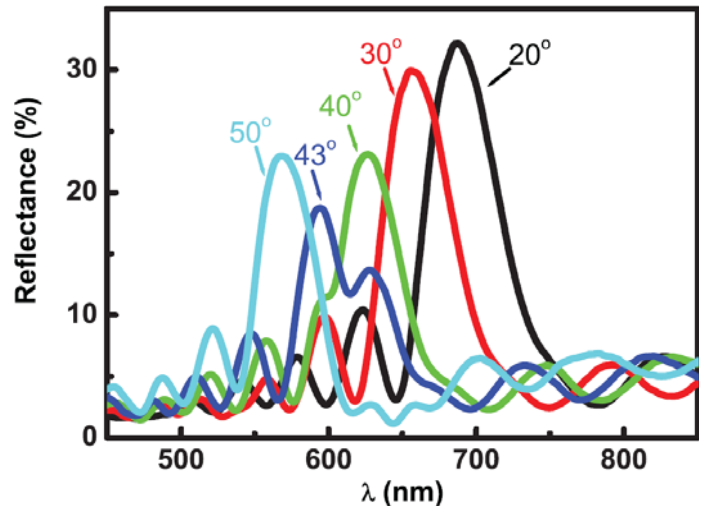

**Figure 2.** (Specular reflection spectra of the first opal (9 layers) with sputtered silica for incidence angles between 20° and 50°.)

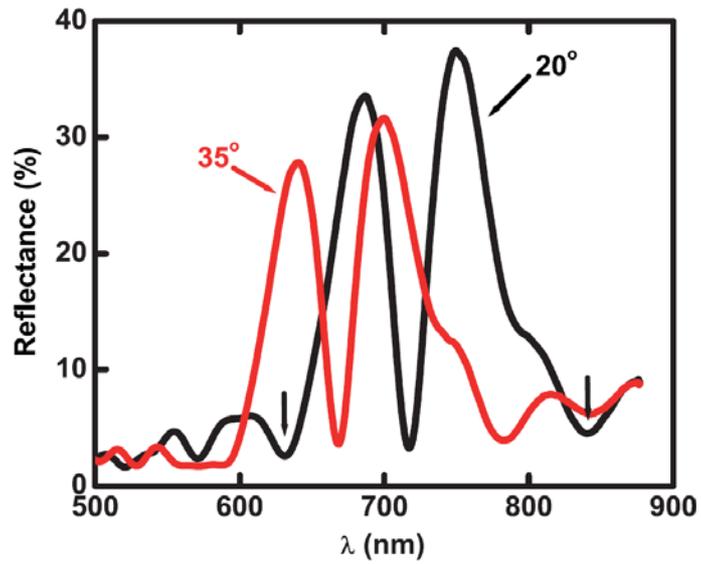

**Figure 3.** (Experimental reflection spectra of the 123 nm-defect thickness sample for 20° (black line) and 35° (red line) incidence angles.)

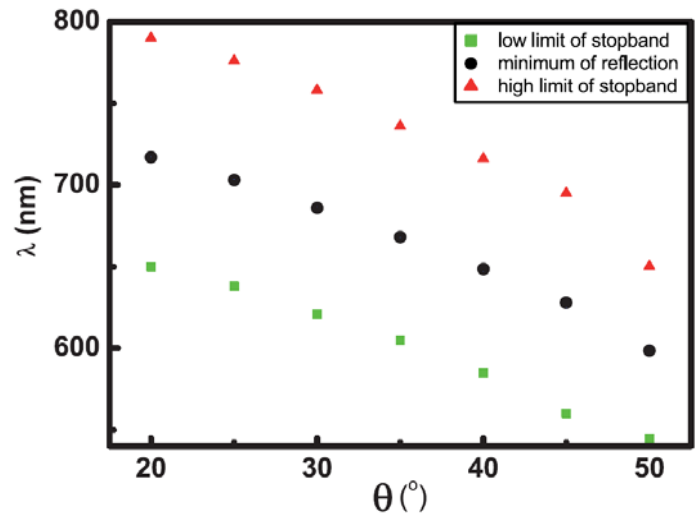

**Figure 4.** (Position of the reflection minima versus the incidence angle (circles). The triangles correspond to the limit of the stopband for high wavelengths, the squares to the limit of the stopband for low wavelengths both deduced from the experimental spectra.)

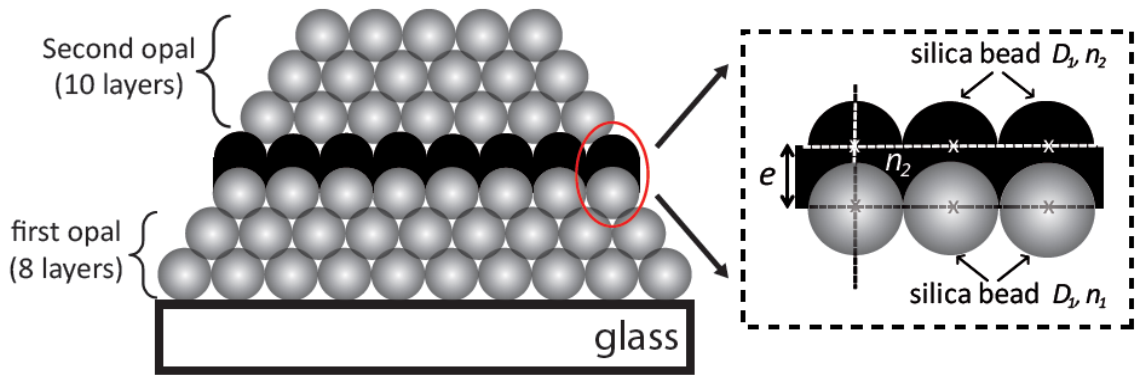

**Figure 5.** (Scheme of a computed structure for FDTD simulations.)

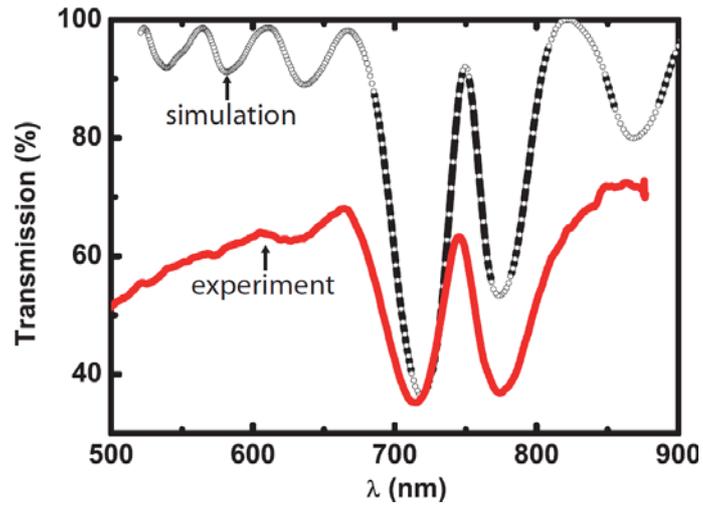

**Figure 6.** (Experimental (red line) and calculated (black line) 0°-transmission spectra of 123 nm-defect thickness sample.)

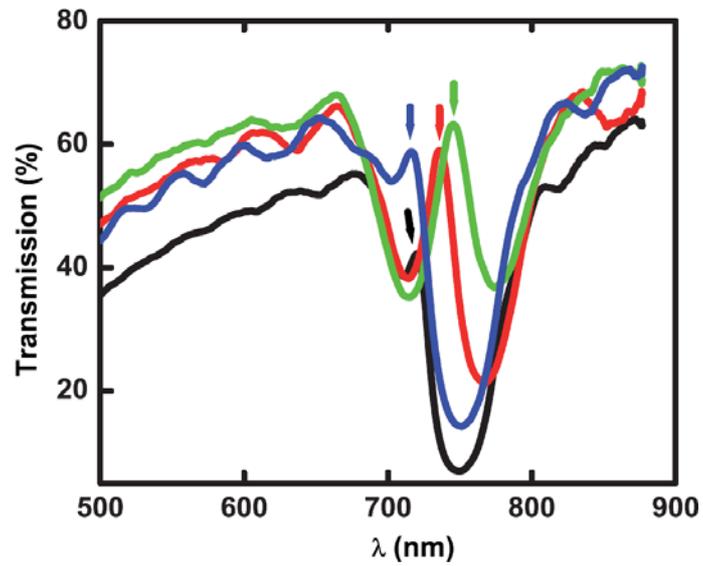

**Figure 7.** (Experimental transmission spectra for 41 nm (blue); 87 nm (red); 123 nm (green); 273 nm (black) of sputtered silica.)

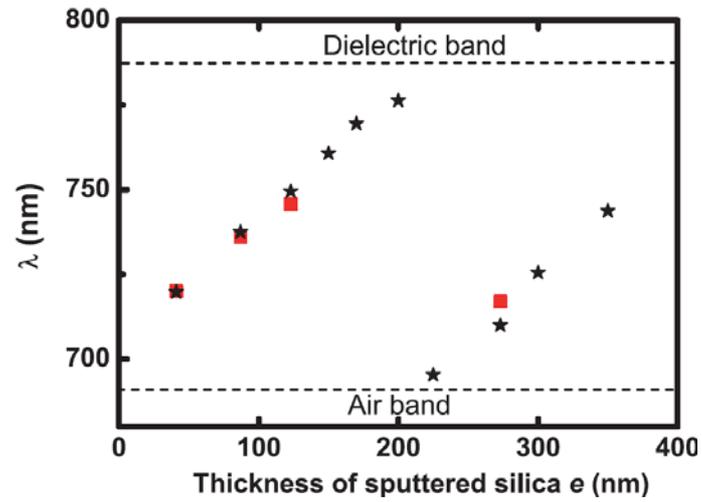

**Figure 8.** (Maximum wavelength of the defect mode for different thicknesses of sputtered silica: simulations (star black point), measurements (rectangle red point). The dashed horizontal lines correspond to the edges of the stopband determined from the experimental spectra (**Figure 7**).)